\documentstyle[12pt]{article}
\newlength{\dinwidth}
\newlength{\dinmargin}
\newlength{\extraspace}
\newlength{\extraspaces}
\newcommand{\be}{\begin{equation}
\addtolength{\abovedisplayskip}{\extraspaces}
\addtolength{\belowdisplayskip}{\extraspaces}
\addtolength{\abovedisplayshortskip}{\extraspace}
\addtolength{\belowdisplayshortskip}{\extraspace}}
\newcommand{\ee}{\end{equation}}
\newcommand{\bdm}{\begin{displaymath}
\addtolength{\abovedisplayskip}{\extraspaces}
\addtolength{\belowdisplayskip}{\extraspaces}
\addtolength{\abovedisplayshortskip}{\extraspace}
\addtolength{\belowdisplayshortskip}{\extraspace}}
\newcommand{\edm}{\end{displaymath}}
\renewcommand{\thefootnote}{\fnsymbol{footnote}}
\def\simlt{\mathrel{\lower2.5pt\vbox{\lineskip=0pt\baselineskip=0pt
           \hbox{$<$}\hbox{$\sim$}}}}
\def\simgt{\mathrel{\lower2.5pt\vbox{\lineskip=0pt\baselineskip=0pt
           \hbox{$>$}\hbox{$\sim$}}}}
\catcode`@=11
\newcount\@tempcntc
\def\@citex[#1]#2{\if@filesw\immediate\write\@auxout{\string\citation{#2}}\fi
  \@tempcnta\z@\@tempcntb\m@ne\def\@citea{}\@cite{\@for\@citeb:=#2\do
    {\@ifundefined
       {b@\@citeb}{\@citeo\@tempcntb\m@ne\@citea\def\@citea{,}{\bf ?}\@warning
       {Citation `\@citeb' on page \thepage \space undefined}}%
    {\setbox\z@\hbox{\global\@tempcntc0\csname b@\@citeb\endcsname\relax}%
     \ifnum\@tempcntc=\z@ \@citeo\@tempcntb\m@ne
       \@citea\def\@citea{,}\hbox{\csname b@\@citeb\endcsname}%
     \else
      \advance\@tempcntb\@ne
      \ifnum\@tempcntb=\@tempcntc
      \else\advance\@tempcntb\m@ne\@citeo
      \@tempcnta\@tempcntc\@tempcntb\@tempcntc\fi\fi}}\@citeo}{#1}}
\def\@citeo{\ifnum\@tempcnta>\@tempcntb\else\@citea\def\@citea{,}%
  \ifnum\@tempcnta=\@tempcntb\the\@tempcnta\else
   {\advance\@tempcnta\@ne\ifnum\@tempcnta=\@tempcntb \else \def\@citea{--}\fi
    \advance\@tempcnta\m@ne\the\@tempcnta\@citea\the\@tempcntb}\fi\fi}
\catcode`@=12
\newcommand{\SM}{Standard Model}
\newcommand{\TDM}{Two-Doublet Model}
\newcommand{\tdm}{two-doublet model}
\newcommand{\br}{branching ratio}
\newcommand{\Ma}{M_{A^{0}}}
\newcommand{\Mg}{M_{H^{\pm}}}
\newcommand{\la}{\lambda}
\newcommand{\La}{\Lambda}

\newcommand{\pr}{Phys.\ Rev.\ }
\newcommand{\prp}{Phys.\ Rep.\ }
\newcommand{\prl}{Phys.\ Rev.\ Lett.\ }
\newcommand{\np}{Nucl.\ Phys.\ {\bf B}}
\newcommand{\pl}{Phys.\ Lett.\ {\bf B}}

\newcommand{\cmp}{Comm.\ Math.\ Phys.\ }
\newcommand{\zp}{Z. Phys.\ {\bf C}}

\newcommand{\rmp}{Rev.\ Mod.\ Phys.\ }
\begin{document}
\begin{titlepage}
\begin{flushright}
BU-HEP 94-5\\
hep-ph/9402339\\
February 22, 1994
\end{flushright}
\vspace{24mm}
\begin{center}
\Large{{\bf The Phenomenology of the CP-odd Scalar
in Two-Doublet Models}}
\end{center}
\vspace{5mm}
\begin{center}
Dimitris Kominis\footnote{e-mail address: kominis@budoe.bu.edu}
\\*[3.5mm]
{\normalsize\it Dept. of Physics, Boston University, 590 Commonwealth
Avenue,}\\
{\normalsize\it Boston, MA 02215}
\end{center}
\vspace{2cm}
\thispagestyle{empty}
\begin{abstract}
We examine the phenomenology of the $CP$-odd scalar $A^0$ of
two-Higgs-doublet models. We explore the parameter space determined by
triviality bounds and identify the regions where the $A^0$ can be
detected at the LHC in each of the following modes: the inclusive
two-photon decay mode, the $l^{\pm}\gamma\gamma X$ mode from
$t\overline{t}A^0$ production and the $A^0\rightarrow Zh$
channel with the subsequent decay of the $CP$-even scalar $h$ to two
photons. We find that, while the $l^{\pm}\gamma\gamma X$ mode is of
limited usefulness, the other two modes can give viable signals in
fairly large, and complementary, regions of parameter space.
\end{abstract}
\end{titlepage}
\newpage

\renewcommand{\thefootnote}{\arabic{footnote}}
\setcounter{footnote}{0}
\setcounter{page}{2}
\section{\bf Introduction}
Despite the remarkable experimental success of the \SM\ of electroweak
interactions, there is no evidence, so far, in favor of the minimal
one-doublet scalar sector of the model. It is therefore essential that
alternatives to it be considered. The simplest {\it extension\/} of the
scalar sector of the \SM\ that naturally embodies certain indispensable
features, such as the smallness of $\rho -1$ and the suppression of
flavor-changing neutral currents, while admitting the possibility of new
observable phenomena, is the ($CP$-conserving) two-Higgs-doublet model.

In the Standard one-doublet Model, the Higgs mass is constrained to lie
between 57 and 800 GeV approximately. The lower limit is set by direct
searches at LEP \cite{lep}, while the upper bound comes from theoretical
considerations such as the triviality of theories with fundamental
scalars \cite{trivt,dn,trivl}.
The numerous phenomenological investigations of the
one-doublet \SM\ \cite{smphen}
indicate that it may be possible to explore this
entire range at future colliders such as LEP-II and the LHC. So we may
ask about the potential of these colliders to detect alternative models.
For a general two-Higgs-doublet model,
a first step towards answering this question was
made in \cite{me}, where the bounds on the parameters of the model that
result from triviality considerations were derived. These
bounds were obtained in the context of perturbation theory and so their
validity is of a rather qualitative nature. To be definite,
we will focus our interest on the region of parameter space
defined by these triviality bounds. However,
the general discussion we offer can give a fair idea of how to extend our
results beyond this region.

In the analysis of ref. \cite{me}, following the spirit of \cite{dn},
the term `triviality' was employed in
the following sense: The running scalar self-couplings of
the \TDM\ are expected, at least in perturbation theory, to develop a
Landau pole at a finite momentum scale. Consequently, the theory can
only stand as an effective low-energy theory valid up to some finite
cutoff $\Lambda$, beyond which new phenomena emerge. The calculation of
physical quantities in the effective theory will thus be accurate up to
terms of order $p_i^2/\La ^2, M_j^2/\La ^2$, where $p_i$ are typical
momenta of the process under consideration and $M_j$ are the masses of
the particles of the theory. In ref. \cite{me}, the effective theory was
defined to be valid if all masses satisfied
\be
\frac{M_j}{\La} \leq \frac{1}{2\pi}.
\ee
Thus, given a set of parameters of the model, that is, masses and
couplings, a cutoff was defined by
\be
\La = 2\pi\,\max_{j} {M_j}
\ee
and the following consistency requirements were made: (i) No coupling
should develop a Landau pole at a scale less than
$\La$, and (ii) the effective potential
should be stable for all field values less than $\La$. From these two
conditions follow the `triviality bounds' reported in \cite{me} and used
in the present study. 

It is thus our intention to find out how much of the parameter space
allowed by triviality can be explored at present and future colliders.
In this paper we concentrate on the
phenomenology of the $CP$-odd scalar (which we denote $A^0$) of the
\TDM\ at the LHC, and
identify those regions of parameter space where direct observation of
this particle is possible.

We examine three of the most promising signals for the $A^{0}$.
First, we discuss the inclusive two-photon channel. As is the case for
the Standard Higgs, a light $A^{0}$ decays predominantly into
$b\overline{b}$ pairs and thus its detection at a hadron collider has to
rely on rare modes, such as the two-photon mode. This signal usually
consists of a large number of events and, furthermore, it may be useful
even beyond the so-called `intermediate mass' range, because the $A^{0}$
does not couple to a pair of weak gauge bosons at tree level.
On the other hand,
it is well known that there are very large backgrounds to this signal,
whose successful containment places severe requirements on the design of
the detector \cite{gem}. Consequently,
even a large event rate may not constitute a
clean signal. For this reason, we also examine the process in which
the $A^{0}$ is produced in
association with a $t\overline{t}$ pair and decays into two photons
while one of the $t/\overline{t}$ is tagged by its leptonic decay
\cite{marciano,gunplb}.
The lepton tag ensures a much cleaner signal, but the event rate
is so small as to render this mode only marginally useful. Finally, we
turn to the discussion of the decay $A^{0}\rightarrow Zh$, where $h$ is a
$CP$-even scalar, with the subsequent decay of $h$ in two photons, and
demonstrate that it gives a very clear signal
in a region of parameter space where none of the two other modes
considered can be useful. To our best knowledge, this process has not
been considered before in this context. Our results indicate that it may
provide an excellent way of directly observing the CP-odd scalar $A^0$.

Phenomenological studies of two-Higgs-doublet models have been carried
out mostly in the context of the Minimal Supersymmetric Model (MSSM)
\cite{kz,mssm,gunorr,ghkao,bcps,hawaii};
if the supersymmetric partners are heavy enough, then the
low-energy spectrum and the formal structure of the couplings of this
model are
exactly the same as in the more general non-supersymmetric case.
Supersymmetry, however, imposes relations among the various masses and
couplings, which reduce the number of independent parameters and allow
for more definite predictions. Moreover, the parameter space usually
investigated in studies of the MSSM is motivated by supersymmetric
unification, while in the present study it is defined by the
above-mentioned triviality bounds and, as a result, it is significantly
different. It might be thought that the MSSM should really be contained
in the more general \tdm\ as a special case. The models studied here,
however, employ an exact discrete symmetry (see Section 2 and ref.~\cite{gw})
intended to suppress flavor-changing neutral currents at tree level; if
this symmetry is allowed to be softly broken, as is the case in the
MSSM, most triviality bounds no longer hold \cite{me}. In this case the
parameter space of the MSSM is indeed a subset of that of the more
general \tdm.

The larger number and different range of the parameters of the \TDM\ may
result in a markedly different phenomenology from that of the MSSM.
A striking example is the
$A^{0}\rightarrow Zh\rightarrow Z\gamma\gamma$ decay sequence,
described in detail in Section 7. We
will further comment on the similarities and differences between the
supersymmetric and non-supersymmetric two-doublet models in the course
of the discussion of our results.

In the next section we review the \TDM, emphasizing the features
most relevant to phenomenology. In Section 3 we describe the production
mechanisms of the $CP$-odd Higgs $A^{0}$ at the LHC. In Section 4 we
discuss the main decay channels and present branching ratios of the
$A^{0}$ in order to motivate our choice of
the particular signals we subsequently examine.
Section 5 concerns the inclusive two-photon channel, while in
Section 6 we study the $l\gamma\gamma X$ signal from $t\overline{t}A^{0}$
production. Section 7 is devoted to the discussion
of the decay $A^{0}\rightarrow Zh$. The results we present in Sections 5-7
consist in event-rate
and significance contours in appropriate sections of parameter space,
for integrated luminosities of 10 fb$^{-1}$ and 100 fb$^{-1}$, which
correspond to a low- and a high-luminosity option at the LHC. We have
used a center-of-mass energy of $\sqrt{s}=16$~TeV in all calculations.
Finally, Section 8 contains our conclusions.

\section{The two-doublet model}
The scalar sector contains two electroweak doublets
$\Phi_{1},\;\Phi_{2}$, both of hypercharge $Y=1$. In order to eliminate
flavor-changing neutral currents at tree level, one has to impose a
discrete symmetry \cite{gw}. There are many ways of doing this, the two
most often discussed being the following\footnote{The notation here has been
changed from that of ref. \cite{me} to facilitate comparison with
most studies of the MSSM. Contact with \cite{me} can be made by the
following substitutions: $\Phi_{1} \leftrightarrow \Phi_{2}$,
Model~I $\leftrightarrow$ Model~II.  In terms of the angles $\alpha$ and
$\beta$ introduced later in this Section,
this has the effect \mbox{$\alpha \leftrightarrow
-\alpha$}, \mbox{$\beta \leftrightarrow \pi/2-\beta$.}} \cite{hhg,sher}:
\be
\begin{array}{lcl}
\bullet\ \, {\rm Model\ I} & : &
\Phi_{1}\rightarrow -\Phi_{1} \\*[2.5mm]
\bullet\ \, {\rm Model\ II} & : & \Phi_{1}\rightarrow -\Phi_{1}
\mbox{\hspace{0.5cm} ;\hspace{0.5cm}}  d_{Ri}\rightarrow -d_{Ri},
\mbox{\hspace{0.5cm}} e_{Ri}\rightarrow-e_{Ri} \mbox{\hspace{32mm}}
\end{array}
\label{discrete}
\ee
($d_{Ri}$ ($i=1,2,3$) are the right-handed negatively charged quarks and
$e_{Ri}$ the right-handed charged leptons.) The Lagrangian is
\bdm
{\cal L} ={\cal L}_{kin}+{\cal L}_{Y}-V
\edm
where ${\cal L}_{kin}$ contains all the covariant derivative terms, $V$ is the
scalar potential and
${\cal L}_{Y}$ contains the fermion-scalar interactions. The form of
the latter is the following:
\begin{itemize}
\item Model I
\be
{\cal L}_{Y}=g_{ij}^{(u)}\overline{q}_{Li}\Phi_{2}^{c}u_{Rj}+
g_{ij}^{(d)}\overline{q}_{Li}\Phi_{2}d_{Rj}+g_{ij}^{(l)}\overline{l}_{Li}
\Phi_{2}e_{Rj}\;+\;{\rm h.c.}
\ee
\item Model II
\be
{\cal L}_{Y}=g_{ij}^{(u)}\overline{q}_{Li}\Phi_{2}^{c}u_{Rj}+
g_{ij}^{(d)}\overline{q}_{Li}\Phi_{1}d_{Rj}+g_{ij}^{(l)}\overline{l}_{Li}
\Phi_{1}e_{Rj}\;+\;{\rm h.c.}
\ee
\end{itemize}
where $q_{Li}$ are the left-handed quark doublets, $u_{Ri}$ are the
up-type right-handed quarks and $l_{Li}$ are the left-handed lepton
doublets. The matrix $g^{(l)}$ is understood to be diagonal.

Thus in Model~I only $\Phi_{2}$ couples to fermions, while in Model~II
$\Phi_{1}$ gives mass to down-type quarks and leptons and $\Phi_{2}$ to
up-type quarks. Because of the different fermion couplings, the two
models have different phenomenology. We will be presenting results for each of
the two models\footnote{It should be noted that the MSSM is of type II
only.}.

The discrete symmetry (\ref{discrete}) also limits the number of scalar
self-couplings. The scalar potential is
\begin{eqnarray}
V & = & \mu_{1}^{2}\,\Phi_{1}^{\dag}\Phi_{1}+
\mu_{2}^{2}\,\Phi_{2}^{\dag}\Phi_{2}
+\la_{1}\,(\Phi_{1}^{\dag}\Phi_{1})^{2}+\la_{2}(\Phi_{2}^{\dag}\Phi_{2})^{2}
+\la_{3}\,(\Phi_{1}^{\dag}\Phi_{1})(\Phi_{2}^{\dag}\Phi_{2}) \nonumber \\
  & & +\la_{4}\,(\Phi_{1}^{\dag}\Phi_{2})(\Phi_{2}^{\dag}\Phi_{1})
+\frac{1}{2}\la_{5}\,[(\Phi_{1}^{\dag}\Phi_{2})^{2}
+(\Phi_{2}^{\dag}\Phi_{1})^{2}] \label{potential}
\end{eqnarray}
\noindent
Provided certain inequalities among the scalar self-couplings are satisfied
\cite{ineq}, the doublets $\Phi_{1}, \Phi_{2}$ acquire vacuum
expectation values of the following form:
\be
\langle\Phi_{1}\rangle=\frac{1}{\sqrt{2}} \left( \begin{array}{c}
0 \\ v_{1}
\end{array}  \right) \mbox{\hspace{1.1cm}  \hspace{1.1cm}}
\langle\Phi_{2}\rangle=\frac{1}{\sqrt{2}} \left( \begin{array}{c}
0 \\ v_{2}
\end{array}  \right)
\ee
where $v_{1}, v_{2}$ are real and
\be
v_{1}^{2}+v_{2}^{2}\equiv v^{2}=(246\ {\rm GeV})^{2}.
\label{vev}
\ee
Note that $CP$ is a good symmetry of the scalar sector
and so we may assign definite $CP$ quantum numbers to all neutral scalar
states. At higher loop level, interactions with the fermions will
invalidate this statement and mixings between $CP$-odd and $CP$-even
states will appear due to the $CP$-violating phases of the
Kobayashi-Maskawa matrix. These mixings, however, are small, and we will
therefore neglect them in what follows.

In addition to the Goldstone bosons which become the longitudinal
components of the $W$'s and the $Z$ via the Higgs mechanism, the
spectrum of the scalar sector contains two neutral $CP$-even scalars,
denoted by $h,H$, one neutral $CP$-odd scalar, $A^{0}$, sometimes
referred to as a pseudoscalar due to its $\gamma^{5}$ coupling to
fermions, and two charged complex conjugate states $H^{\pm}$. It is
customary to introduce two mixing angles: \mbox{$\beta$
$(=\arctan (v_{2}/v_{1}))$} rotates the $CP$-odd and the charged scalars into
their mass eigenstates while $\alpha$
\mbox{$(-\pi/2 \!<\!\alpha \! \leq \pi/2)$}
rotates the neutral scalars into their mass eigenstates. There are six
independent parameters in the scalar potential (after fixing $v$ through
(\ref{vev})) which can be taken to be $\alpha, \beta, M_{H^{\pm}},
M_{H}, M_{h}, M_{A^{0}}$. (In contrast, in the MSSM there are only
two.) To these one should add the top quark mass, expected to lie in the
range \cite{top}
\be
100\ {\rm GeV} \;\;\simlt \;\;M_{t} \;\;\simlt \;\;200\ {\rm GeV}
\ee
but otherwise unspecified.

In Table 1 we display the couplings of the neutral scalars to
vector bosons and fermions for Models I and II.
\begin{table}
\begin{center}
\begin{tabular}{|c|c|c|}     \hline
  &  Model~I  &  Model~II \\ \hline
$hVV$ & $\sin (\beta-\alpha)$ & $\sin(\beta-\alpha)$ \\ \hline
$HVV$ & $\cos (\beta-\alpha)$ & $\cos(\beta-\alpha)$ \\ \hline
$A^{0}VV$ & 0 & 0 \\  \hline
$hu\overline{u}$ & $\cos \alpha/\sin \beta$ & $\cos\alpha/\sin\beta$ \\
\hline
$hd\overline{d}$ & $\cos\alpha/\sin\beta$ & $-\sin\alpha/\cos\beta$ \\  \hline
$Hu\overline{u}$ & $\sin\alpha/\sin\beta$ & $\sin\alpha/\sin\beta$  \\  \hline
$Hd\overline{d}$ & $\sin\alpha/\sin\beta$ & $\cos\alpha/\cos\beta$  \\  \hline
$A^{0}u\overline{u}$ & $i\gamma^{5}\cot\beta$ & $i\gamma^{5}\cot\beta$ \\
\hline
$A^{0}d\overline{d}$ & $-i\gamma^{5}\cot\beta$ & $i\gamma^{5}\tan\beta$ \\
\hline
\end{tabular}
\caption{Weak gauge boson and fermion couplings of the neutral scalars of the
\tdm\ relative to the corresponding Standard Higgs couplings.}
\end{center}
\end{table}
The values shown are
relative to the \SM\ Higgs couplings; $V$ denotes collectively the weak
gauge bosons, $u$ the up-type quarks and $d$ the down-type quarks and
leptons. We also display below the couplings of $A^{0}$ to a scalar and
a vector boson, since they are important in the determination of the
partial widths of $A^{0}$.
\be
{\cal L}_{A^{0}SV}=\frac{1}{2}\sqrt{g^{2}+g^{'2}}\,Z_{\mu}\,
[A^{0}\stackrel{\leftrightarrow}{\partial ^{\mu}}
(H\sin (\beta-\alpha)-h\cos (\beta-\alpha))]
-\frac{1}{2}\,g\,W_{\mu}^{+}
(A^{0}\stackrel{\leftrightarrow}{\partial ^{\mu}}
H^{-}) \;+\; {\rm h.c.} \label{azh}
\ee
where $g^{'},g$ are the $U(1)_Y$ and $SU(2)$ gauge couplings
respectively. The full
set of Feynman rules can be found in ref. \cite{hhg}.

Of crucial importance to the phenomenology of the $CP$-odd Higgs $A^{0}$
is the absence of a tree level coupling between it and a pair of weak gauge
bosons. This deprives us of a very clean signal as well as of an
important production mechanism for masses of a few hundred GeV. It is
also evident from Table 1 that the couplings of the Higgs bosons to
fermion-antifermion pairs can be significantly enhanced or suppressed
over those of the Standard Higgs depending on the values of $\beta$ and
$\alpha$ and the choice of Model. Similar suppressions can occur in the
couplings of the $A^{0}$ to the other neutral scalars (see eq. (\ref{azh})).
Unlike the situation in the MSSM, these couplings are independent of the
scalar masses. This freedom has important consequences for
phenomenology, as will be demonstrated in later sections.

The strongest experimental constraints on the \tdm\ arise from the
recent CLEO bound on $BR(b\rightarrow s\gamma)$ \cite{cleo} and from LEP
data on the $Z\rightarrow b\overline{b}$ decay width
\cite{zbbbar}. These measurements impose a combined
bound on $\beta$ and $M_{H^{\pm}}$ \cite{buras,park}. There are uncertainties
\cite{buras} about
the precise value of these bounds, and we will not include them in our
analysis. From refs. \cite{buras,park}, however, it is clear
that light charged scalars ($M_{H^{\pm}}
\simlt 50\ {\rm GeV}$) and low $\beta$ ($\beta \simlt 15-20 ^{o}$) are
strongly disfavoured, with the restrictions becoming tighter for larger
top quark masses.

\section{Production mechanisms}
As was emphasized earlier, the coupling of the $A^{0}$ to intermediate
vector bosons is induced only at loop level and is therefore small. So,
in contrast to the case of the Standard Higgs, production of the $A^{0}$
via vector boson fusion or in association with a $W$ or a $Z$ is
insignificant. Thus, at a hadron collider, we have to rely on the
following mechanisms:
\bdm
\begin{array}{lrcl}
{\rm (a)\;\; Gluon\ fusion} & gg & \rightarrow & A^{0} \\
{\rm (b)\;\; Associated\ } b\overline{b} A^{0}\ {\rm production} \hspace{1cm} &
q\overline{q}, gg & \rightarrow & b\overline{b} A^{0} \\
{\rm (c)\;\; Associated\ } t\overline{t} A^{0}\ {\rm production} &
q\overline{q}, gg & \rightarrow & t\overline{t} A^{0}
\end{array}
\edm
\noindent
The cross sections for processes (a)--(c) are identical to the
corresponding rates in the MSSM, since they only depend on $\Ma,
M_{t}$ and $\tan \beta$. We discuss them here for
completeness and for future reference. Our results are in agreement with
ref. \cite{ghkao}.

\subsection{Gluon fusion}
This is usually the dominant mode. The reaction proceeds via a quark
loop \cite{tri}. The contribution of the top quark is generally the most
significant, but, in Model II and for large $\beta$, the bottom quark
contribution can become equally important or even dominate, especially
for small $\Ma$. We have taken into account leading QCD corrections as
follows: in ref. \cite{kaufman} the QCD corrections to the production of
a pseudoscalar have been calculated in the limit of a very heavy quark
flowing in the loop. The major part of these corrections can be cast
into the form
\be
\sigma = \sigma_{0} \left[1+(6+\pi^{2})\frac{\alpha_{s}(M_{A^{0}})}{\pi}
\right]
\label{qcd}
\ee
in analogy to the case of a $CP$-even Higgs \cite{kz}. $\sigma_{0}$ and
$\sigma$ are the lowest-order and QCD-corrected cross-sections
respectively. For our purpose, this is a valid approximation only if
$\Ma < 2M_{t}$ and only insofar as the bottom loop contribution, to
which this correction does not apply, is negligible compared to that of
the top quark. With hindsight, we may say that the second qualification
is not worrisome: in the regions where the bottom quark
contribution is significant, namely large $\beta$, small $\Ma$, the
rates for the signals we will be interested in are small anyway. This may
not be so for $\beta$ very close to 90$^o$, but triviality bounds almost
always exclude such values of $\beta$.
The QCD corrections to arbitrary mass pseudoscalar production from
gluon fusion have not been computed. Consequently, we shall only
employ (\ref{qcd}) if $M_{A^{0}}<2M_{t}$.

Figures 1a, 1b show the production cross-section of the $CP$-odd Higgs
through gluon fusion as a function of its mass for Models I and II
respectively. The effect of the bottom quark loop for large $\beta$ in
Model II is evident. The top quark mass was taken to be 150~GeV.
The curves for different $M_{t}$ are similar; they always peak at
$M_{A^{0}}=2\,M_{t}$, as long as $\beta$ does not take on values
extremely close to 90$^{o}$ in Model II.
The value at the peak is lower for higher $M_{t}$.
Here and throughout this study we employed the EHLQ structure
functions, Set 2, with $\Lambda_{\overline{MS}}=290$ MeV \cite{ehlq}.

\subsection{Associated $b\overline{b}A^{0}$ production}
This mechanism can be important for large $\beta$ in Model II when the
$b\overline{b}A^{0}$ coupling becomes strong.
For $M_{A^{0}} \gg M_{b}$, this process can be approximated
\cite{dicusw} by
$b\overline{b}$ fusion using the bottom quark distribution functions
within the proton. The $b\overline{b}$ fusion cross-section is
proportional to the partial width $\Gamma(A^{0}\rightarrow
b\overline{b})$. QCD corrections to the latter have been included in our
calculation as we explain in Section 3. No other QCD corrections to
this process have been estimated.

In Figures 2a, 2b we show the production cross-section of $A^{0}$ from
$b\overline{b}$ fusion as a function of the $A^{0}$ mass for Models I
and II. It can be seen that, in the case of Model I, this cross-section
never exceeds the gluon fusion cross-section. In contrast, this
mechanism dominates in Model II for very large values of $\beta$.

\subsection{Associated $t\overline{t}A^{0}$ production}
This process always gives a smaller rate than gluon fusion. It is
interesting, however, in its own right, because it is possible to obtain
a relatively clear signal if one can tag on a lepton from the decay of the
top or antitop. This will be further discussed in Section~6. We have
computed this cross-section using the spinor techniques of
ref. \cite{kleisstir}. QCD corrections
have not appeared in the literature and
are not included. In Figures 3a, 3b we display the cross-section for the
process $pp\rightarrow t\overline{t}A^{0}$ as a function of the
pseudoscalar mass and for the set of values of $\beta$ used in the
previous graphs. The top quark mass is 120~GeV and 180~GeV
respectively. There is only a slight drop in rate with increasing top
quark mass, as the smaller parton luminosities at high mass are
compensated by the enhanced Yukawa coupling. There is no distinction
between Models I and II in this case because the $A^{0}t\overline{t}$
coupling is the same in both Models (see Table 1). We also show the
corresponding curves for the \SM\ Higgs. In all calculations
we have included the contributions
of both the $gg$ and $q\overline{q}$ initial states. The latter is
negligible for the $CP$-odd scalar but amounts to \mbox{20-25\%} of the
gluon-initiated
reaction for a Standard Higgs lighter than about 200 GeV (see also
\cite{summpre}). We notice that the curves for the $CP$-odd Higgs are
much flatter than those for the Standard Higgs. For $\beta=45^{o}$,
the $t\overline{t}A^{0}$ production rate for $A^{0}$ masses
less than about 150 GeV
is significantly lower than the corresponding rate
for the Standard Higgs. As we will see in Section 6, this has unpleasant
phenomenological implications.

In Figures 4a, 4b we show the total production cross-section for the
$A^{0}$, the sum, that is, of the three processes discussed above.
Results are shown for the two Models and for the same values of $\beta$
used in previous figures. A top quark mass of 150 GeV was assumed. We
again display the corresponding cross-section for the Standard Higgs.
(Gauge boson fusion has not been included in the calculation of the
latter.) The conclusions to be drawn from
these graphs are the following: (i) except for very large
values of $\beta$ in Model II, the production cross-section increases
with decreasing $\beta$; (ii) the two Models diverge
only at large $\beta$; (iii) for $\beta\simlt 50^{o}$ the cross-section
is larger than that of the Standard Higgs;
(iv) the cross-section is generally peaked at
$M_{A^{0}}=2M_{t}$.  Variation of $M_{t}$ shifts
the peak accordingly and lowers the value at the peak somewhat. The
other parameters of the \tdm\ do not affect the production rate of the
$CP$-odd scalar.

\section{Decays and Branching Ratios}
In this Section we review the major decay modes and the corresponding
\br s of the $CP$-odd Higgs in order to identify the signals that can be
of phenomenological interest.
In the absence of a tree level coupling to
the $W$'s and the $Z$, the $A^{0}$ decays mainly to fermion pairs or to
states containing other scalars. Loop-induced decays can also be large,
like the two-gluon mode, or phenomenologically interesting, like the
two-photon decay\footnote{See \cite{ghkao}, however, for an
investigation of the mode \mbox{$A^{0}\rightarrow ZZ$} in the MSSM.}.
The main channels are the following:
\be
A^{0}\rightarrow
b\overline{b},\;c\overline{c},\;t\overline{t},\;\tau^+\tau^-,
\;W^{\pm}H^{\mp},\;Zh,\;ZH,\;gg,\;\gamma \gamma,\;Z\gamma.
\label{decays}
\ee
All decay widths have been calculated to lowest non-trivial order.
Analytic formulae for the last three, loop-induced, decays can be found
in the literature \cite{loopdecay,hhg,ghkao}.
QCD corrections have been included where they
are most significant, namely:

(i) the $q\overline{q}$ decays ($q=b,c$): Leading QCD effects have been
accounted for by using, in the expression for the Yukawa coupling,
the running quark mass at the scale of the mass
of the $CP$-odd Higgs \cite{kks};

(ii) the $\gamma \gamma$ and $Z\gamma$ decays through quark loops:
Corrections have been calculated in ref. \cite{djouadiga} for the
$\gamma \gamma$ decay. The leading contribution can be incorporated by
using the running quark mass at scale $M_{A^{0}}$. The remaining QCD
corrections are mostly within 10\% and have not been included. We have also
used running masses in the computation of the decay to $Z\gamma$;

(iii) the $gg$ decay: Corrections to this decay have only
been computed for the case of a $CP$-even Higgs \cite{djouadigl}.
In analogy to that calculation, we used running quark masses in the
loops, but did not include any further corrections.

The \br s of the $CP$-odd scalar are complicated functions of the seven
free parameters of the model and hence impossible to describe
completely. We will illustrate some of the most prominent features by
means of a few examples.

Fig. 5 shows the \br s of the $A^{0}$ as a function of its mass for a
case where the other scalars, $h,H$ and $H^{\pm}$,
are heavier than the top quark. The set of
values of the parameters chosen (referred to as Set A) is the following:
\be
\begin{array}{llll}
{\rm Set\ A:}\mbox{\hspace{0.5cm}} & M_{H}=400\; {\rm GeV},\; &
M_{h}=260\; {\rm GeV},\; & \Mg=350\; {\rm GeV}, \\ &
M_{t}\;\,=120\; {\rm GeV},\; & \alpha=30^{o}. &
\end{array}
\label{seta}
\ee
Branching ratios are shown for two values of $\beta$ ($30^{o}$ and
$60^{o}$) and for Models~I and II. Of the decay channels listed in
(\ref{decays}) we do not show here the \br s to
$c\overline{c}, gg, ZH$ and $Z\gamma$. The first two are
phenomenologically uninteresting at hadron colliders. The $ZH$ channel
has a behaviour qualitatively similar to that of the $Zh$ mode, while
the \br\ to $Z\gamma$, a channel of potential phenomenological
importance, is generally very small. We point out the following
features: (i) For $M_{A^{0}}<2M_{t}$, the dominant decay mode is
$b\overline{b}$, unsuitable for detection of a Higgs at a hadron
collider.\footnote{$\beta=30^o$ in Model~II illustrates the case of
the $gg$ \br\ (not shown) becoming dominant just below the top
threshold.}
Thus, the $\gamma \gamma$ or $\tau^{+}\tau^{-}$ decays have to be
employed. (ii) Whereas the \br\ to two photons in Model~I is insensitive
to $\beta$, in Model~II it is a decreasing function of $\beta$, at least
for moderate $\beta$ values. This is easy to understand; the two-photon
width is governed by the $A^{0}t\overline{t}$ coupling, because of the
top loop that mediates the decay, and
so goes like $\cot^{2}\beta$ (see Table~1),
whilst the full width is essentially
saturated by the $b\overline{b}$ mode ($\sim \tan^{2}\beta$). As a
result, the $\gamma \gamma$ \br\ has an approximate $\cot ^{4} \beta$
dependence. In Model~I, however, the $A^{0}$ couplings to
$b\overline{b}$ and $t\overline{t}$ are both
proportional to $\cot\beta$ and
the $\beta$-dependence drops out. (iii) The channels that contain
scalars ($Zh$, $W^{\pm}H^{\mp}$) become dominant, or at least compete
with the $t\overline{t}$ mode, as soon as they open. (iv) Both the
$\tau^{+}\tau^{-}$ and the $\gamma \gamma$ \br s fall abruptly once the
$t\overline{t}$ channel becomes available. In the region between the
$t\overline{t}$ and the scalar particle thresholds, detection of the
$A^{0}$ seems dubious.

With Fig.~6 we wish to illustrate the dependence of the \br s on the
angle $\alpha$ instead. In what we call Set~B in this figure, the
parameters have the following values:
\be
\begin{array}{llll}
{\rm Set\ B:}\mbox{\hspace{0.5cm}} & M_{H}=400\; {\rm GeV},\; &
M_{h}=100\; {\rm GeV},\; & \Mg=350\; {\rm GeV}, \\ &
M_{t}\;\,=180\; {\rm GeV},\; & \alpha=30^{o}, &
\end{array}
\label{setb}
\ee
whereas in `Set~C' they are
\be
\begin{array}{llll}
{\rm Set\ C:}\mbox{\hspace{0.5cm}} & M_{H}=400\; {\rm GeV},\; &
M_{h}=100\; {\rm GeV},\; & \Mg=350\; {\rm GeV}, \\ &
M_{t}\;\,=180\; {\rm GeV},\; & \alpha=-45^{o}. &
\end{array}
\label{setc}
\ee
$\beta$ is fixed at $40^{o}$ in all graphs of this figure. This is an
example of the $Zh$ channel opening up before the $t\overline{t}$ does.
The main points we want to emphasize here are the following: (i) $\gamma
\gamma$ and $\tau^{+}\tau^{-}$ are again the only possibilities for
detection of a light $A^{0}$; (ii) the $\alpha$-dependence comes from
the $A^{0}Zh$ coupling (see eq.~(\ref{azh})) which is proportional to $\cos
(\beta-\alpha)$. In Set~B this coupling is not suppressed and the $Zh$
mode is dominant even after the $t\overline{t}$ channel becomes
available. In contrast, in Set~C, the $Zh$ \br\ is sizeable only before
the $t\overline{t}$ threshold. What is also to be noticed in this case
is that the opening of the $Zh$ channel does not cause any significant
drop to the value of the two-photon \br . As a result, the $\gamma
\gamma$ decay remains a promising mode for $A^{0}$ masses as large as
$2M_{t}$. It should be noted that no such suppression is present in the
$A^0W^{\pm}H^{\mp}$ coupling (see eq.~(\ref{azh})). Consequently, the
two-photon \br\ necessarily drops once the $W^{\pm}H^{\mp}$ channel is
open, independently of the value of $\alpha$.

Although the particular shapes of these curves change for different
values of the \TDM\ parameters, the general features described above
persist. We are thus led to examine the observability of the $CP$-odd
Higgs in the two-photon mode (as the most promising alternative for a
light $A^{0}$, but potentially useful even for an $A^{0}$ of mass as
large as $2M_{t}$) and in the $Zh$ mode (which, unless it is suppressed,
covers a range of higher masses $M_{A^{0}}$). We have not further
considered the $\tau^{+}\tau^{-}$ mode\footnote{See \cite{kz} for a
discussion of this channel in the MSSM. The
$\tau^{+}\tau^{-}$ \br\ for a light $A^0$
is of the order of 10\% in both the MSSM and the
\TDM; consequently, the findings of \cite{kz} should apply here too.}.

\section{The inclusive two-photon channel}
This mode has been extensively studied in the literature
for the $CP$-odd scalar of the MSSM \cite{kz,gunorr,bcps,hawaii}.
Since the
parameter space under investigation here is qualitatively rather
different, as explained in the Introduction, an independent study is
necessary. We shall briefly compare our findings to those for the
MSSM, to the extent that this is possible, at the end of
this Section.

To assess the observability of a Higgs signal in this mode a reliable
calculation of the reducible and irreducible backgrounds is required.
This, in turn, presupposes a good understanding of the detector
parameters and a full Monte Carlo simulation. We did not make such an
effort here, but only tried to approximate a realistic situation as much
as possible, within the framework of simple parton-level calculations.
Our procedure of estimating the backgrounds, which are the same as in
the \SM, is sketched below.

At lowest order, the direct two-photon background arises from the Born
process
\be
q\overline{q} \rightarrow \gamma \gamma
\label{born}
\ee
\noindent
The gluon fusion process
\be
gg \rightarrow \gamma \gamma,
\label{box}
\ee
through the `box' diagram, gives a contribution of the same order of
magnitude as (\ref{born}). We calculated the cross-sections for
processes (\ref{born}) and (\ref{box}) using the following kinematical
cuts to simulate the detector acceptance:
\be
|\eta^{\gamma}|<2.5 \mbox{\hspace{2cm}};\mbox{\hspace{2cm}}
p_{T}^{\gamma}>20\; {\rm GeV}
\label{ggcuts}
\ee
In the calculation of the `box' diagram we used five light flavours and
a top quark of 150~GeV. Variation of the top quark mass in the range
100~GeV $\!<\!M_{t}\!<\!$ 200~GeV produces negligible change in the
cross-section. A next-to-leading order calculation of the di-photon
background \cite{nlo,aurenche} shows that bremsstrahlung processes,
where one or more photons
are radiated off a jet, largely dominate over the lowest order ones.
However, suitable isolation cuts \cite{aurenche} can reduce
this higher-order background to about half the sum of the Born
and `box' contributions. We therefore multiplied the
calculated cross-section of processes (\ref{born}) and (\ref{box})
by a `K-factor' of 1.5 in
order to `convert' to the next-to-leading-order result.

The background from events where jets are faking photons can be
significant. We assumed (following \cite{seez,cms}) that it will be
possible to achieve such a jet rejection efficiency as to reduce these
backgrounds to well below the prompt di-photon level. In
fact, even so, the jet backgrounds can still be as large as 25\% of the
physical $\gamma \gamma$ background \cite{seez}, depending on the
two-photon invariant mass. We adopted this figure as our estimate of the jet
background that remains after all isolation cuts, applying it to all
invariant masses. Finally, we assumed
a 90\% reconstruction efficiency for each photon as well
as a further 7\% loss, from isolation cuts, due to pileup, if the high
luminosity option is used \cite{seez}.
To obtain the number of background events, we used a mass bin of
$\Delta M_{\gamma \gamma} =3\%
M_{\gamma \gamma}$. This is probably a conservative choice, given the
energy resolution capability quoted in refs. \cite{seez,cms}.
However, precise vertex identification is a very difficult task
\cite{gem}, especially at high luminosity, and it is uncertain how
accurately the vertex will be localized \cite{seez}. A conservative
assumption on the vertex identification capability leads to an energy
smearing such that our choice of $\Delta M_{\gamma \gamma}$
then corresponds, approximately, to a $2\sigma$ bin width (see
\cite{seez}).
In the cases where the signal was broader than this mass bin, we
summed the background over the width of the resonance.

The signal event rate depends on all six parameters of the  scalar
sector as well as on the top quark mass and the
choice of Model. The discussion of Sections 3 and 4
has shown that the most important parameter, besides
$\Ma$, is $\beta$, by virtue of its effect on both the production
rate and the two-photon decay width of the $CP$-odd Higgs. We will
therefore present results fixing the other parameters at first.
Nonetheless,  we will discuss the effect of varying them and illustrate
it by means of some examples.

In Section 4 it was observed that the widths of the $t\overline{t}$ and
scalar decay modes are generally large, so that the \br\ to two photons
is, most of the times, greatly reduced once these channels are open.
Figures~7 and~8 reflect this behaviour. They display the production
cross-section times the two-photon \br\ of the $A^{0}$ as a function of
its mass for various values of $\beta$. Fig.~7 corresponds to Set~A of
the parameters (see (\ref{seta})), while Fig.~8 corresponds to Set~B
(eq.~(\ref{setb})). We
also plot, in each case, the same quantity for the Standard Higgs. In
these calculations we included QCD corrections, as explained in
Sections~3 and~4, but not cuts or efficiencies. The curves do not
therefore represent the actual signal cross-sections, but, rather, they
are shown for purposes of comparison. As expected from the discussion of
the preceding sections, the rate increases with $\Ma$ until the
$t\overline{t}$ or $Zh$ threshold is met; beyond this, the rate falls due
to the sharp decrease of the $\gamma \gamma$ \br. In Fig.~8, the
cross-section rises again right before $\Ma=2M_{t}$, reflecting
the increased production rate at this mass.
(The only exception to
this general pattern is the case of large $\beta$ in Model~II; there,
the quark loop diagrams responsible for both the $A^{0}$ production from
gluon fusion and the decay to two photons are dominated by the bottom,
rather than the top,
quark and so do not possess a maximum at $\Ma=2M_{t}$.)
Variation of $M_{t}, M_{H^{\pm}}, M_{H}, M_{h}$
may change the order in which
the various thresholds appear, but has little effect otherwise.
The dependence on the angle $\alpha$ is generally mild, except for its
influence on the couplings of $A^{0}$ to $Zh$ and $ZH$, discussed at
length in Section~4. Figure~9 corresponds to Set~C of parameters (see
(\ref{setc})), an example where the suppression of the $A^{0}Zh$
coupling is particularly strong.

Comparison with the Standard Higgs, in Figures~7-9, shows that in the
so-called intermediate-mass range, the rates for the $A^{0}$ are
smaller, unless $\beta\simlt 40^{o}$. On the other hand, they can
remain substantial for a much wider range, reflecting the fact that the
$WW$ and $ZZ$ decay modes are effectively absent for the $CP$-odd
scalar.

In order to determine the statistical significance of a signal
in this mode, we computed its rate using the cuts (\ref{ggcuts})
and the efficiencies
quoted in the discussion of the backgrounds. We only accepted signals
that consist of at least 40 events and defined the significance by the
ratio $S/\sqrt{B}$. In Figs.~10-12, we plot $5\sigma$ contours for
integrated luminosities of
10 fb$^{-1}$ and 100 fb$^{-1}$, and for each of
Sets~A,~B and~C. The contour corresponding to 10 fb$^{-1}$ can
also be regarded as a $15\sigma$ contour\footnote{Here and in subsequent
Sections we effectively scale the number of signal and background events
with the integrated luminosity (except for the small pileup effect
mentioned earlier). It is not absolutely clear, however, whether the
assumed detector efficiency can be maintained in a high luminosity
environment \cite{gem}.} for an integrated luminosity of
100 fb$^{-1}$; it thus serves also as an indication of the variation of
the significance in the parameter space depicted. As explained in the
Introduction, we restrict attention to the region of parameter space
allowed by triviality constraints; this region is shown explicitly in
Figs.~10-12. The significance plots show
somewhat more quantitatively the behaviour expected on the basis of the
results displayed in Figs.~7-9: For Sets~A and~B the opening of the
$t\overline{t}$ and $Zh$ channels respectively, makes the detection of
$A^{0}$ virtually impossible beyond these thresholds, despite the small
window around $\Ma=2M_{t}$ (Set~B). On the other hand, the
suppression of the strength of the $A^{0}Zh$ coupling leads to rather
spectacular results for Set~C. It is worth noting that this is not a
case of fine-tuning: any $\alpha$ in the range
$-60^{o}\simlt\alpha\simlt -30^{o}$ would have more or less a similar
effect.

Our comment on the dependence of the curves of Figs.~7-9 on the values
of the other parameters of the model can help one visualize the
variation of these significance contours
with $M_{t}, M_{H^{\pm}}, M_{H}, M_{h}$ and
$\alpha$: we expect the masses to play a role only insofar as
they set the various thresholds, and $\alpha$ to smoothly effect the
transition from a situation like that of Fig.~11 to one like
that of Fig.~12, as it
gradually alters the factor of $\cos(\beta-\alpha)$ governing the $Zh$
\br\ of the $A^{0}$. (The latter is relevant only if the $Zh$
threshold comes before the $t\overline{t}$ or $W^{\pm}H^{\mp}$ ones.)
However, there is a different way in which all these parameters may affect
our conclusions, and this is by changing the triviality bounds. The
allowed area may be shifted to lower or higher $\beta$, expanded or
shrunk according to the precise values of the parameters chosen. This
effect won't be too drastic if the scalars and top quark are not
very light or very heavy; however, an example is shown in Fig.~13, where
the region in which the $A^{0}$ can be observed in this mode is rather
narrow. We note the values chosen for this example for future reference:
\be
\begin{array}{llll}
{\rm Set\ D:}\mbox{\hspace{0.5cm}} & M_{H}=400\; {\rm GeV},\; &
M_{h}=100\; {\rm GeV}, & \Mg=350\; {\rm GeV}, \\
 & M_{t}\;\,=180\; {\rm GeV}, & \alpha=-75^o. &
\end{array}
\label{setd}
\ee

In the MSSM, values of $\tan\beta>1$ are preferred \cite{bcps}. In this
case, since $\alpha$ is not an independent parameter,
$\cos(\beta-\alpha)$ is constrained to be small, especially for large
$\Ma$; thus the $A^{0}Zh$ coupling is suppressed, leading to a
situation similar to that of Fig.~12b. However, the bias on $\tan\beta$
restricts us to a region of low production cross-section making this
mode appear less valuable overall than in the \TDM.

As a conclusion, we may say that there is a substantial region of
parameter space where the inclusive two-photon mode provides a clear
signal for the $A^{0}$. This region extends from the lowest allowed
$\beta$ to $\beta \sim 55-60^{o}$, with the signal fading as $\beta$
increases, and from $\Ma \sim 40$ GeV to the first threshold
of a `strong' decay (by which we mean
$t\overline{t}, W^{\pm}H^{\mp}$ or $Zh$ if
not suppressed), with the signal improving with larger $\Ma$.
This is true for both
Model~I and Model~II. From the examples considered it can be seen
that Model~II is slightly less favourable to high $\beta$ values than
Model~I, but otherwise exhibits similar behaviour. These
results, however, were based on rather
optimistic assumptions about the ability of
the detector to successfully reduce the jet backgrounds. This
uncertainty has led many authors to seek alternative methods of
detecting the scalar particles of the \SM\ and its extensions
\cite{marciano,gunplb,kun84}. We
now turn to a method that has recently attracted a lot of
attention.

\section{The $l\gamma \gamma X$ channel}
In the \SM, the Higgs boson can be produced in association with a $W$ or
a $t\overline{t}$ pair through the following processes:
\begin{eqnarray}
q\overline{q} & \rightarrow & W^{\star} \rightarrow WH \label{wh} \\
gg,\; q\overline{q} & \rightarrow & t\overline{t}H \label{tth}
\end{eqnarray}
\noindent
The leptonic decay of the $W$ or the $t/\overline{t}$ gives rise to a
final state that consists of two isolated photons from the Higgs and an
isolated lepton ($e$ or $\mu$). Although the event rate is not high, the
signal is rather clean, at least
at low luminosity \cite{marciano,gunplb,kts,bal,sumplb}.

For the $CP$-odd scalar of the \TDM, (\ref{wh}) gives a negligible rate,
as stated earlier. Consequently, we concentrate on (\ref{tth}). The
final state from this process contains, besides the photons and the
lepton, at least two jets. This can be used to remove most of the
reducible backgrounds \cite{kts,sumplb}, as will be explained below.

The irreducible backgrounds arise from the processes
\begin{eqnarray}
gg & \rightarrow & t\overline{t} \gamma \gamma \label{ggf} \\
q\overline{q} & \rightarrow & t\overline{t} \gamma \gamma \label{qqf}
\end{eqnarray}
\noindent
We did not perform an independent calculation of these backgrounds, but
relied instead on the results of ref.~\cite{gunorr} for (\ref{ggf})
\footnote{These authors
employed the HMRSB parton distribution functions \cite{hmrsb}. We scaled
their results by 1.5 to account for the fact that EHLQ usually yield
higher rates for gluon-initiated processes \cite{gunplb}.}. The cuts
assumed by these authors were the following:
\be
\begin{array}{lcr}
|\eta^{l,\gamma}| < 2.5 \mbox{\hspace{2cm}} & , &\mbox{\hspace{1cm}}
p_{T}^{l,\gamma} > 20\; {\rm GeV}, \\
\Delta R\,(\gamma_{1},\gamma_{2}) > 0.4 & , & \Delta R\,(l,\gamma) > 0.4.
\end{array}
\label{lggcut}
\ee
\noindent
Insisting that at least two jets be detectable in the final state
results in a loss of rate that varies from 25\% for $M_{t}=100$~GeV
to 4\% for $M_{t}=160$~GeV \cite{sumplb}. (In \cite{sumplb}
a jet was deemed detectable if it passed the same cuts as a photon or a
lepton.) The background (\ref{qqf}) was calculated in
ref.~\cite{summpre} and was
found to amount to 50\% of (\ref{ggf}) at $M_{\gamma\gamma}=100$~GeV
climbing to 120\% at $M_{\gamma\gamma}=180$~GeV. We assumed
these values in our estimate of the background.

A potentially serious background could arise from the process $gg
\rightarrow b\overline{b}\gamma\gamma$, where a $B$ meson decays
semileptonically. Since, however, the matrix element for this process is
peaked when the $b$ quarks are emitted along the beam direction, the
cuts imposed on jets reduce this background to negligible levels
\cite{sumplb}.
Other reducible backgrounds arise from events involving $t\overline{t}$
production in which one or two jets in the final state are misidentified
as photons. In
\cite{gunorr}, the background from $t\overline{t}g$ production was
calculated and found negligible assuming a $\gamma$-jet rejection factor
of $5\times 10^{-4}$. A rough estimate of $t\overline{t}\gamma$
production was also reported and the resulting background declared to be
small compared to the $t\overline{t}\gamma \gamma$ background, though
not negligible. Another possible source of background is $t\overline{t}$
production, where both photons are faked by jets from top decays. We
assumed that these backgrounds can all be kept below the irreducible
$t\overline{t}\gamma\gamma$ level. However, a detailed study is
necessary in order to assess their exact magnitude at the LHC.

To obtain the background event number we used the reconstruction
efficiency for photons and leptons (65\% overall) quoted in \cite{seez}
and assumed, as before, a 3\% resolution for the invariant mass of the
photons. The background is a decreasing function of the top quark mass,
due to the smaller parton luminosities at higher energies. Moreover, it
depends indirectly on the mass of the charged Higgs: if the decay mode
$t\rightarrow H^{+}b\,$ is kinematically allowed, then the leptonic ($e$
and $\mu$) \br\ of the top (or antitop) is reduced from the \SM\ value of
2/9 because $H^{+}$ decays to electrons or muons are very rare. This
remark applies, of course, also to the signal cross-section, to the
discussion of which we now turn.

Fig.~14 shows the cross-section times \br\ for the process
$pp\rightarrow t\overline{t}A^{0} \rightarrow l\gamma\gamma X$, as a
function of $\Ma$ for various values of $\beta$. The other parameters
are chosen to belong to Set~A (see (\ref{seta})).
The corresponding curve for the Standard
Higgs is also shown. Cuts were not included in this calculation. The
variation of the curves with the top quark mass is illustrated in
Fig.~15, where $M_{t}=180$~GeV while the remaining \TDM\ parameters
are as in Set~A. We shall refer to this choice as `Set~E'. We note that,
due to the weak dependence of the production cross-section on the mass
of the top,
the rates are not too different for Sets~A and~E as long as $\Ma <
2M_{t}$. The production cross-section does not depend on parameters
other than $\Ma, M_{t}$ and $\beta$, while the variation of the \br\
with the other parameters has been discussed in Section 4. Consequently,
we expect, as before, a sudden drop of the signal rate when the
$t\overline{t}$ or scalar thresholds appear, unless again there is a
suppression of the $A^{0}Zh$ coupling. What is noteworthy in Figs.~14-15
is that for Higgs masses of less than about 160 GeV and for a large range
of values of $\beta$ the rate is well below its \SM\ value, especially
in Model~I. Considering the fact that a Standard Higgs of this mass can
only give a few tens of events \cite{gunplb,kts,bal},
we recognize that the prospects
of detecting $A^{0}$ in this mode are rather poor. In compensation,
however, the mass range where this mode could be -- even marginally --
helpful may be much wider.

In Figs.~16-17 we display significance contours for Sets~A and~E. Here the
calculation of the signal was carried out including the cuts
(\ref{lggcut}), reconstruction efficiencies and losses from insisting
that at least two jets be detected in the final state. The fraction of
signal lost due to this last requirement is the same as for the
background (see ref.~\cite{sumplb}).
The kinematical cuts result in a signal loss
that varies from 75\% at $\Ma=30$ GeV to about 30\% at $\Ma=300$ GeV
(with a very soft dependence on the top quark mass).
We required that a signal candidate should
consist of more than 10 events. For low numbers of events, the
formula for the significance used in Section 5 is not very accurate
because it is based on Gaussian statistics. Since the
production of `events' is a Poisson process, the probability that
an apparent signal is in fact due to background
fluctuations is \cite{gem}
\be
P=\sum_{n=S+B}^{\infty} P_n(B)
\ee
where $S,B$ are the numbers of expected signal and background events
respectively and $P_n(B)$ is the probability that $n$ background events
occur, given by the Poisson distribution
\be
P_n(B)=B^n e^{-B} /n! \;\; .
\ee
(Since $S+B$ is not an integer in general, an interpolation between the
two integers closest to it is employed.) The probability $P$ can then be
converted into a significance.

The contour for an integrated luminosity of 10 fb$^{-1}$ is absent from
Figs.~16-17; at such low luminosity
the LHC cannot detect the $CP$-odd scalar of the \tdm\ in this mode at the
5$\sigma$ confidence level. For Set~E, Model~I, even a luminosity of
100 fb$^{-1}$ is inadequate.
The main problem is not the background, which
is small, especially for large $M_{t}$, but rather the size of the
signal. A reduced isolation radius of $\Delta R =0.2$ has the effect of
boosting the signal by 20\%. A similar answer was found for the
background in ref. \cite{kts}; an understanding of the effect of this
change in $\Delta R$ on the reducible backgrounds is needed before
deciding whether this procedure can enhance the significance of the
signal.

The chances of observing the $A^{0}$ would be improved if the value of
$\beta$ were lower, since then the production would be raised. In
Fig.~18 we show an example where low $\beta$ values are allowed by
triviality. (This can be achieved, roughly speaking, when $|\alpha|$ is
small, the top quark is light and the scalar masses moderate.) For this
example, we chose the following values:
\be
\begin{array}{llll}
{\rm Set\ F:}\mbox{\hspace{0.5cm}} & M_{H}=350\; {\rm GeV},\; &
M_{h}=150\; {\rm GeV},\; & \Mg=350\; {\rm GeV}, \\ &
M_{t}\;\,=120\; {\rm GeV},\; & \alpha=0^{o}. &
\end{array}
\label{setf}
\ee
We notice
that for $\beta \simlt 30^{o}$ a highly significant signal is possible,
at least for Model~II.
Recall, though, our earlier comment that such low $\beta$ values are in
danger of being already ruled out by experiment. (Both the measurement
of $BR(b\rightarrow s\gamma)$ and that of the $Z\rightarrow b\bar{b}$
decay width set lower bounds on $\beta$; for charged Higgs masses of a
few hundred GeV the bound from the
latter is the strongest \cite{buras}.) If this example
represents the `corner' of parameter space where the prospects of
detection of the $A^{0}$ in this mode are optimal, Set~D (see (\ref{setd}))
of parameter values is such that no point in the ($\Ma \,,\,\beta$) plane
gives a 5$\sigma$ signal in either luminosity.

Summarizing, we remark that the $l\gamma\gamma X$ mode may give a clear
signal in only a narrow range of parameter space, due to the low
$A^{0}t\overline{t}$ production rate, which, for masses below 150~GeV
approximately, is considerably smaller than the corresponding rate for
the Standard Higgs. Consequently, this mode was found to be useful only
if $\beta \simlt 35^{o}$, a range that may easily be disallowed by
triviality (or experiment). If the integrated luminosity at the LHC is
10~fb$^{-1}$, then the $A^{0}$ can be observed in this mode only in
exceptional cases. Comparison with the inclusive two-photon mode reveals
the latter, under the optimistic assumptions listed in Section~5, to be
a more promising option. The $l\gamma \gamma X$ mode can be a useful
independent check, but enough events can only be accumulated if $\beta$
is relatively small and the high luminosity option at the LHC is used.

\section{The $Z\gamma \gamma$ channel}
In Section 4 we observed that the branching fraction of the $A^{0}$ to
the states containing other scalars ($Zh,\;ZH,\;W^{\pm}H^{\mp}$) is
generally large, and suggested that these modes may prove valuable in
the search for the $CP$-odd Higgs, particularly for large $\Ma$ where the
two-photon mode cannot be relied on. In this section we examine in
detail the process
\be
A^{0} \rightarrow Zh \rightarrow l^{+}l^{-}\gamma \gamma
\label{atozl}
\ee
where $l$ stands for an electron or a muon. A similar discussion applies
to the mode $A^{0}\rightarrow ZH$. For definiteness, we shall keep $H$
heavy. Allowing for the possibility that $H$ is light enough to have an
appreciable \br\ to two photons is straightforward to implement, but
would not add to our understanding of the phenomenology of the $A^{0}$.

The direct background to (\ref{atozl}) comes primarily from
the process
\be
q\overline{q} \rightarrow Z\gamma \gamma
\label{qqtozgg}
\ee
\noindent
It was found to be small, as we show below, so we did not consider
necessary to perform the calculation of gluon fusion into $Z\gamma
\gamma$ (through the `pentagon' diagram) which could in principle be
comparable. There is another process that can interfere with
(\ref{qqtozgg}), namely the standard production of a $CP$-even Higgs in
association with a $Z$:
\be
q\overline{q} \rightarrow Z^{\star} \rightarrow Zh\; (\rightarrow
Z\gamma\gamma).
\ee
This was also found to be very small, so we will neglect it in what
follows.
The calculation of (\ref{qqtozgg}) was done using spinor techniques
\cite{kleisstir}. The cross-section is singular when the photons are emitted
along
the beam axis or are soft. The following cuts were employed to exclude
such photons and simulate a realistic situation:
\be
\begin{array}{lcr}
|\eta^{l,\gamma}| < 2.5 \mbox{\hspace{2cm}} & , &\mbox{\hspace{1cm}}
p_{T}^{l,\gamma} > 20\; {\rm GeV}, \\
\Delta R\,(\gamma_{1},\gamma_{2}) > 0.4 & , & \Delta R\,(l_{1},l_2) > 0.4, \\
\Delta R\,(l,\gamma) > 0.4 & . &
\end{array}
\label{zggcut}
\ee
\noindent
The isolation cut for photons and leptons has only a 5\% effect on this
calculation. We also included reconstruction
efficiencies of 90\% for each lepton and 85\%
for each photon after isolation cuts (see ref.~\cite{seez}).

Fig.~19a shows a plot of the number of $Z\gamma\gamma$ background
events per GeV per LHC year (at a luminosity of
$10^{34}\,{\rm cm}^{-2}\,{\rm s}^{-1}$) as a function of the $\gamma
\gamma$ invariant mass. For an invariant mass in the range
40~GeV $\simlt\! M_{\gamma\gamma}\!\simlt $ 160~GeV (the interesting mass
range for a Higgs $h$ to decay to two photons), a 3\% detector
resolution will yield $\sim$ 2 events of background. A similar
conclusion is reached by looking at Fig.~19b, which plots events per GeV
per LHC year versus the total final state invariant mass. For typical
values of parameters that result in a signal of several events, the width of
$A^{0}$ is a few GeV. The background in such a bin is about 1
event. (The $A^{0}$ can be considerably broader at high mass, but there
the $M_{Z\gamma\gamma}$ distribution dies out.) The $h$ and $A^{0}$
resonances will appear as bumps in the $M_{\gamma\gamma}$ and
$M_{Z\gamma\gamma}$ distributions respectively, and it is the {\em same}
events that will constitute both bumps. This is not the case with the
background, however. The background events with a two-photon invariant
mass $M_{\gamma\gamma}$ within 3\% of $M_{h}$ will have a broad
distribution in $M_{Z\gamma\gamma}$ and will not, in general, be
concentrated in the vicinity of the $A^{0}$ resonance. This observation
enables us to effectively eliminate the background. For example, Higgs
masses of $M_{h}=\,$100~GeV, $\Ma=\,$300~GeV may typically give a signal
of a few tens of events. Of the $\sim \! 2$ background events that have
$M_{\gamma\gamma}$ inside a 3\% window around 100~GeV, less than 0.1
lie within 5~GeV of $\Ma$. Hence it becomes plain that the background
that simultaneously emulates both $h$ and $A^{0}$ is truly negligible.

QCD corrections to process (\ref{qqtozgg}) have not been calculated. We
also did not estimate the size of reducible backgrounds from jet events
($Z\gamma j,\;Zjj$). It is assumed that, with the efficiencies and
rejection capabilities mentioned in Section 5 \cite{seez}, these
backgrounds will be brought to below the level of the direct
$Z\gamma\gamma$ production. But in view of the arguments of the
preceding paragraph, even
if the reducible backgrounds were to dominate by as much as an order of
magnitude over the irreducible $Z\gamma\gamma$ production, they would
still pose no significant problems.

It is evident, therefore, that the observability of the $A^{0}$ in this
mode depends only on whether a sufficient number of signal events can be
accumulated. The magnitude of the light neutral Higgs \br\ to two photons,
$BR(h\rightarrow \gamma\gamma)$, is thus of crucial importance. This \br\
has a complicated dependence on the various masses and angles. We will
indicate very roughly what this dependence is so as to make the pattern
of our results more intelligible.

A Higgs $h$ in the intermediate-mass range 40~GeV $\simlt M_{h} \simlt
160$~GeV, decays mostly to a $b\overline{b}$ pair. Therefore, to a good
approximation,
\be
BR(h\rightarrow \gamma\gamma) \approx \frac{\Gamma(h\rightarrow
\gamma\gamma)}{\Gamma(h\rightarrow b\overline{b})}
\ee
{}From the $hb\overline{b}$ couplings displayed in Table~1, it follows
that
\be
\Gamma (h\rightarrow b\overline{b}) \sim \left\{ \begin{array}{ll}
\cos ^{2}\alpha /\sin ^{2}\beta & \mbox{ (Model~I)} \\
\sin ^{2}\alpha /\cos ^{2}\beta & \mbox{ (Model~II)}
\end{array}
\right.
\ee
\noindent
The partial width $\Gamma(h\rightarrow \gamma\gamma)$ is more
complicated. The light neutral
Higgs $h$ decays to two photons through a loop in which a $W$, an
$H^{+}$ or a charged fermion flows. For moderate values of $\beta$ the
contribution of the $W$ is the largest, although, in some cases, the
top quark loop is of comparable magnitude. The $hWW$ coupling (see Table~1)
is proportional to $\sin (\beta-\alpha)$ so we may write, very roughly,
\be
BR(h\rightarrow \gamma\gamma) \sim \left \{ \begin{array}{ll}
\sin ^{2}(\beta-\alpha) \sin ^{2}\beta /\cos ^{2}\alpha &
\mbox{(Model~I)} \\
\sin ^{2}(\beta-\alpha) \cos ^{2}\beta /\sin ^{2}\alpha &
\mbox{(Model~II)}
\end{array} \right.
\label{hbr}
\ee
\noindent
provided none of the sines and cosines that appear in these formulae
are too small. Another feature of the $\gamma \gamma$ \br\ of the
scalar $h$ is that it becomes largest for $M_{h} \approx $ 120-130~GeV
and has a very soft dependence on $\Mg$ and $M_{t}$.

Our prime motivation for considering this decay mode was to explore
regions of parameter space where the two-photon channel gives a signal
too faint to observe. In Fig.~20 we show $Z\gamma\gamma$ event
contours for Set~D of parameters (see (\ref{setd})). More than 10 events
are expected for points inside these contours. The
solid curve (corresponding to an integrated luminosity of $L=10\,{\rm
fb}^{-1}$) can also be interpreted as a 100-event
contour for $L=100\,{\rm fb}^{-1}$. The
\br\ for the $Z$ decay into electrons or muons, as well as the
kinematical cuts (\ref{zggcut}) and the lepton and photon reconstruction
efficiencies, were included in the
calculation of the signal.
QCD corrections to the decays of $h$ were
taken into account in a way similar to that described in Section 4 for
the $A^0$ \footnote{In addition, the known corrections to the two-gluon
mode \cite{djouadigl} were also included.}.
The large value of $|\alpha|$ (see
(\ref{setd})) results in a huge rate in the case of Model~I, where, at
$L=$~100~fb$^{-1}$, a clear signal is obtained for almost the entire
parameter space depicted. For Model~II the results are more modest, but
comparison with Fig.~13 shows that, even in this case, the
$Z\gamma\gamma$ signal is observable in a region where the two-photon
mode can hardly be helpful. Choosing $|\alpha|$ to be small would
instead produce a large rate in Model~II. This is illustrated in
Fig.~21. The only difference relative to Set~D is the value of $\alpha$
which in this example is $\alpha=0^o$. We shall refer to this choice as
`Set~G'. Here too, the inclusive two-photon mode fails to give an
observable signal for $A^{0}$ masses beyond the threshold for
the $Zh$ decay. (We omit plots for this, however.)

In Section 5 we remarked that the $\gamma\gamma$ signal is weak for
large $\Ma$, when other channels open up, and for large $\beta$, when
the production rate (as well as the two-photon \br\ in Model~II) is
small. From (\ref{hbr}) it follows, however, that, depending on $\alpha$
and the choice of Model, the light $CP$-even Higgs \br\ to two photons
may be enhanced at large $\beta$ and perhaps even compensate for the
loss in production rate. An example where this happens is given in
Fig.~22. The parameters belong to Set~C, which represents the case where the
$A^{0}Zh$ coupling is weak enough to allow a sizeable $\gamma\gamma$
\br\ to persist up to the top threshold (see Fig.~12). Indeed, the $Zh$
mode can't give a good signal beyond this threshold either, but, at high
luminosity, it
covers some of the large-$\beta$ regions undetectable by the two-photon
mode, while being able to confirm the latter for smaller $\beta$ and
$M_{Z}+M_{h} \!<\!\Ma\!<\! 2M_{t}$.

In order to convey an idea on the dependence of our results on $\alpha$,
we present in Fig.~23 similar 10-event contours in the
($\alpha$,$\beta$) space, now keeping $\Ma$ fixed:
\be
\begin{array}{llll}
{\rm Set\ H:}\mbox{\hspace{0.5cm}} & M_{H}\,=400\; {\rm GeV},\; &
M_{h}=100\; {\rm GeV}, & M_{H^{\pm}}=350\; {\rm GeV}, \\
 & \Ma=280\; {\rm GeV}, & M_{t}\,=150\; {\rm GeV}.&
\end{array}
\label{seth}
\ee
The precise value of $\Ma$ has no effect on these curves, as long as it
is below the $W^{\pm}H^{\mp}$ or $t\overline{t}$ threshold. The
different behaviour of the two Models for large and small $|\alpha|$ is
evident from the figure. We observe that large numbers of events are
expected in a considerable region of parameter space. The only area left
unexplored corresponds to $\alpha>0$ and large~$\beta$.

The situation is not as rosy if $\Ma\!>\!2M_{t}$ or $\Ma\!>\!M_{W}+\Mg$,
in which case the \br\ to $Zh$ is smaller. A clear signal of a few tens
of events is then guaranteed in Model~I (resp. Model~II) only for large
(resp. small) $|\alpha|$.

In all figures shown the value $M_{h}=100$~GeV was assumed. We remarked
earlier that the \br\ of $h$ to two photons increases with $M_{h}$ until
approximately $M_{h}\approx 120$~GeV and then falls again. The
corresponding $Z\gamma\gamma$ event rates would of course follow the
same trend.

In the MSSM the decay sequence $A^0\rightarrow Zh \rightarrow
l^+l^-\gamma \gamma$
discussed in this section has not been
considered, presumably because it gives too low a rate.
We have
mentioned, in Section 5, that the bias on $\tan \beta$ tends to suppress
the $A^{0}Zh$ coupling in the kinematical region where the decay
$A^{0}\rightarrow Zh$ is possible. Consequently, the \br\ of this decay
is only sizeable when $\tan \beta$ is not very large and
$M_{Z}+M_{h}\!<\Ma\! <\!2M_{t}$ \cite{kz}.
However, in this range, the \br\ of $h$
to two photons does not exceed $10^{-3}$
(except for large $M_{t}$) \cite{kz}.
In contrast, in the context of the non-supersymmetric \TDM, it can reach
the level of a few percent (and even higher in exceptional cases). The
different behaviour can be traced to the fact that $\alpha$ is an
independent parameter in the \TDM\ and may assume values such as to
enhance this \br. The decay $A^{0} \rightarrow Zh$ in the MSSM has
received some consideration in conjunction with the subsequent
$\tau^{+}\tau^{-}$ \cite{AZh} or $b\overline{b}$ \cite{ghkao}
decays of the light
Higgs $h$, with some encouraging results in the former case. The
$\tau^{+}\tau^{-}$ \br\ of an intermediate-mass $h$ in the \TDM, as well
as in the MSSM, is about 10\%, almost independently of $\alpha$ and
$\beta$. Consequently we may expect this mode to be helpful also in the
\TDM; in particular, it may enable us to explore the region for
$\alpha\!>\!0$ where the $Z\gamma\gamma$ signal is weak. A
detailed study of the signal and backgrounds at the LHC is needed in
order to decide this issue.

To summarize, we have found that the process $A^{0}\rightarrow Zh
\rightarrow l^{+}l^{-}\gamma\gamma $ provides an excellent way of
simultaneously detecting the $A^{0}$ and the light $CP$-even scalar $h$,
provided the latter is in the intermediate mass-region 40~GeV $\simlt
M_{h} \simlt$ 160~GeV. This is true even at a yearly luminosity of 10
fb$^{-1}$ at the LHC. Tens or hundreds of signal events can be obtained
over a negligible background in a fairly large region of parameter
space, especially when $|\alpha|$ is large (in Model~I) or small (in
Model~II). This mode only fares poorly when $\alpha$ is positive and
moderate and $\beta$ large, or when the $t\overline{t}$ or
$W^{\pm}H^{\mp}$ channels are available for the decay of $A^{0}$.
If the latter is true, detection is harder but not impossible.

The utility of the $Zh$ channel extends also to larger
masses for the $CP$-even Higgs $h$. If $M_h \simgt 160$~GeV, then $h$
can be detected in its `gold-plated' decay mode $h\rightarrow ZZ\;
{\rm or}\; ZZ^{\star} \rightarrow l^+l^-l^+l^-$, with $l=e,\mu$. We have
estimated that the process $A^0\rightarrow Zh$ followed by the above
decay sequence may yield several events provided $\Ma\!<\!2M_t$.
The last requirement shows that
the kinematical region where this six-lepton signal can be observed is
rather narrow, unless the top quark is relatively heavy
($M_t \simgt 150$~GeV). Nevertheless, this may be the only way of
detecting the $CP$-odd scalar in this region of parameter space.

\section{Conclusions}
In this paper we examined the phenomenology of the $CP$-odd scalar $A^0$
of a two-Higgs-doublet model with an exact discrete symmetry.
We explored the parameter
space defined by the triviality bounds of ref.~\cite{me} in
order to identify the regions where the $A^0$ can be detected at the
LHC. We studied three different signals of the $A^0$. The inclusive
two-photon mode can provide an observable signal if $\beta \simlt
$~55-60$^o$ and provided the decays of the $A^0$ to
$t\overline{t},\;W^{\pm}H^{\mp},\;Zh$ are kinematically forbidden or (in
the case of $Zh$) suppressed. The lower end of the $A^0$ mass range
where detection in this mode is possible depends on $\beta$ and is
raised as $\beta$ increases. Models with small $\beta$ lead to larger
signals because the production rate of $A^0$ is effectively
proportional to $\cot ^2 \beta$, at least
for moderate values of $\beta$. This is
also true for the associated $t\overline{t}A^0$ production, which was
found capable of giving a detectable signal in the
$l^{\pm}\gamma\gamma X$ mode if $\beta \simlt $~35-40$^o$ (and for roughly
the same range of $\Ma$ as in the inclusive two-photon case). However,
this range is mostly disallowed by triviality bounds (as well as
disfavoured by current experimental data). If the integrated luminosity
at the LHC is 10~fb$^{-1}$, this mode can only be helpful in exceptional
cases. In contrast, the process $A^0 \rightarrow Zh$, followed by the
two-photon decay of the light $CP$-even Higgs $h$, can provide a very clear
signal for a substantial region of parameter space, as long as $h$ is in
the so-called intermediate-mass range (40~GeV$\;\simlt \!M_h
\!\simlt\;$160~GeV). (For heavier $h$, its decay to four leptons through a
$Z$ pair may give rise to several events if the masses are such that
$M_Z+M_h\!<\!\Ma\! <\!2M_t$.)
This mode is largely
complementary to the $\gamma\gamma$ mode since it covers regions of high
$\Ma$ as well as (in some instances) high $\beta$. Under certain
circumstances, it may also confirm observation of the $A^0$ in the
two-photon channel. Hundreds of events can be accumulated in this mode
if $|\alpha|$ is large in Model~I or small in Model~II.

If $\Ma\!>\!2M_t$ or $\Ma\!>\!M_W+\Mg$, then the \br\ to $Zh$ is reduced
and our ability to detect the $CP$-odd scalar in this mode is
compromised. Although for some values of $\alpha$ and $\beta$ this mode
would still be a prime option for the direct observation of $A^0$, it
may be worthwhile investigating other processes, such as the
$t\overline{t}A^0$ production with the subsequent decay of $A^0$ to
$t\overline{t}$ \cite{ghkao}.
Another region that, according to our conclusions, escapes observation
at the LHC, is the low-$\Ma$, large-$\beta$ region (essentially
regardless of the values of the other parameters). It may be possible to
explore part of this region at LEP-II, particularly since the
disadvantage of insufficient production at
large $\beta$ will not be present there.

It would be interesting to perform similar studies for the other scalars
of the \TDM, in order to determine how much of the parameter space of
such models can be explored at future colliders. Work in this direction
is under progress.

\vspace{0.9cm}
\noindent
{\Large\bf Acknowledgements}

I wish to thank R.~S.~Chivukula for suggesting to me the project of the
investigation of the phenomenology of \tdm s and discussing it at all
its stages, K.~Lane and M.~V.~Ramana for very useful discussions and
M.~Golden, V.~Koulovassilopoulos and M.~V.~Ramana for allowing
me to use some of their FORTRAN routines. This work was supported in
part under NSF contract PHY-9057173 and DOE contract DE-FG02-91ER40676,
and by funds from the Texas National Research Laboratory Commission
under grant RGFY93-278.

\newpage

\noindent
{\Large\bf Figure captions}
\vspace{0.5cm}
\begin{enumerate}
\item
Production cross-section of the $CP$-odd Higgs
$A^0$ through gluon fusion as a function
of its mass for (a) Model~I and (b) Model~II. Results are shown for
$\beta=15^o, 45^o,70^o$ and $88^o$ and for $M_t=150$~GeV.
\item
Production cross-section of the $A^0$ through $b\overline{b}$ fusion as
a function of its mass for (a) Model~I and (b) Model~II and for the same set
of values of $\beta$ as in Fig.~1.
\item
Cross-section for the associated $t\overline{t}A^0$ production as a
function of the $A^0$ mass for (a) $M_t\!=\!120$~GeV and
(b) $M_t\!=\!180$~GeV
and for the same set of values of $\beta$ as in Fig.~1. There is no
distinction between Models~I and II in this case. The
corresponding cross-section for the Standard Higgs is also shown for
comparison.
\item
Total production cross-section of the $A^0$ as a function of its mass
for (a) Model~I and (b) Model~II and for the same set of values of
$\beta$ as in Fig.~1. The corresponding cross-section for the Standard
Higgs (not including gauge boson fusion production) is also shown. The
top quark mass is taken to be 150~GeV.
\item
Branching ratios of the $A^0$ as a function of its mass for the
following choices of parameters: (a) $M_H\!=\!400\; {\rm GeV},\; M_h\!=\!260\;
{\rm GeV},\; \Mg\!=\!350\; {\rm GeV},\; M_t\!=\!120\; {\rm GeV},\;
\alpha\!=\!30^o,\;\beta\!=\!30^o$, Model~I; (b) same for Model~II; (c) same as
(a)
except $\beta\!=\!60^o$; (d) same as (b) except $\beta\!=\!60^o$.
\item
Branching ratios of the $A^0$ as a function of its mass for the
following choices of parameters: (a) $M_H\!=\!400\; {\rm GeV},\; M_h\!=\!100\;
{\rm GeV},\; \Mg\!=\!350\; {\rm GeV},\; M_t\!=\!180\; {\rm GeV},\;
\alpha\!=\!30^o,\;\beta\!=\!40^o$, Model~I; (b) same for Model~II; (c) same as
(a)
except $\alpha\!=\!-45^o$; (d) same as (b) except $\alpha\!=\!-45^o$.
\item
Production cross-section times branching ratio to two photons for the
$A^0$ as a function of its mass for (a) Model~I and (b) Model~II and for
the following set of parameters: $M_H\!=\!400\; {\rm GeV},\; M_h\!=\!260\;
{\rm GeV},\; \Mg\!=\!350\; {\rm GeV},\; M_t\!=\!120\; {\rm GeV},\;
\alpha\!=\!30^o$. Curves are shown for $\beta=15^o, 45^o, 70^o$ and $88^o$
as well as for the Standard Higgs.
\item
Same as Fig.~7 but for the following values of parameters: $M_H\!=\!400\;
{\rm GeV},\; M_h\!=\!100\;
{\rm GeV},\; \Mg\!=\!350\; {\rm GeV},\; M_t\!=\!180\; {\rm GeV},\;
\alpha\!=\!30^o$.
\item
Same as Fig.~7 but for the following values of parameters: $M_H\!=\!400\;
{\rm GeV},\; M_h\!=\!100\;
{\rm GeV},\; \Mg\!=\!350\; {\rm GeV},\; M_t\!=\!180\; {\rm GeV},\;
\alpha\!=\!-45^o$.
\item
Significance contours for the inclusive two-photon signal in the region
of the \mbox{$(\Ma\,,\,\beta)$} plane allowed by triviality for (a) Model~I and
(b)
Model~II. The other parameters are as in Fig.~7. Contours are shown for
integrated luminosities of 10~fb$^{-1}$ and 100~fb$^{-1}$. The
statistical significance is greater than $5\sigma$ in the interior of
these contours.
\item
Same as Fig.~10 but for the set of parameters of Fig.~8.
\item
Same as Fig.~10 but for the set of parameters of Fig.~9.
\item
Same as Fig.~10 but for the following choice of parameters: $M_H\!=\!400\;
{\rm GeV},\; M_h\!=\!100\;
{\rm GeV},\; \Mg\!=\!350\; {\rm GeV},\; M_t\!=\!180\; {\rm GeV},\;
\alpha\!=\!-75^o$.
\item
Associated $t\overline{t}A^0$ production cross-section times \br\ to an
$l^{\pm}\gamma\gamma$ final state as a function of $\Ma$. This \br\ is
the product of the $A^0\rightarrow \gamma\gamma$ branching fraction and
the probability that at least one of the $t$ and $\overline{t}$ decays
leptonically. Results are shown for (a) Model~I and (b) Model~II and for
$\beta=15^o,45^o, 70^o$ and $88^o$. The remaining parameters are as
follows: $M_H\!=\!400\; {\rm GeV},\; M_h\!=\!260\;
{\rm GeV},\; \Mg\!=\!350\; {\rm GeV},\; M_t\!=\!120\; {\rm GeV},\;
\alpha\!=\!30^o$. The corresponding quantity for the Standard Higgs is also
plotted for comparison.
\item
Same as Fig.~14 except now $M_t\!=\!180$~GeV.
\item
Significance contours for the $l\gamma\gamma X$ mode in the region of
the $(\Ma\,,\,\beta)$ plane allowed by triviality for (a) Model~I and (b)
Model~II. The other parameters are as in Fig.~14. No point in the region
depicted can give a $5\sigma$ signal if a yearly luminosity of
10~fb$^{-1}$ is employed.
\item
Same as Fig.~16, but for the set of parameters of Fig.~15.
\item
Same as Fig.~16, but for the following values of parameters: $M_H\!=\!350\;
{\rm GeV},\; M_h\!=\!150\;
{\rm GeV},\; \Mg\!=\!350\; {\rm GeV},\; M_t\!=\!120\; {\rm GeV},\;
\alpha\!=\!0^o$.
\item
Continuum $Z\gamma\gamma$ background events per GeV per 100~fb$^{-1}$ as
a function of (a) the two-photon invariant mass $M_{\gamma\gamma}$ and
(b) the total final state invariant mass $M_{Z\gamma\gamma}$.
\item
10-event contours for the $Z\gamma\gamma$ signal in the region of the
$(\Ma \,,\,\beta)$ plane allowed by triviality for (a) Model~I and (b)
Model~II. The slice of parameter space shown corresponds to the
following set of parameters: $M_H\!=\!400\;
{\rm GeV},\; M_h\!=\!100\;
{\rm GeV},\; \Mg\!=\!350\; {\rm GeV},\; M_t\!=\!180\; {\rm GeV},\;
\alpha\!=\!-75^o$. Contours are shown for luminosities of 10~fb$^{-1}$ and
100~fb$^{-1}$. More than 10 events are expected in the interior of these
contours.
\item
Same as Fig.~20 except now $\alpha\!=\!0^o$.
\item
Same as Fig.~20 except now $\alpha=-45^o$.
\item
10-event contours for the $Z\gamma\gamma$ signal in the region of the
$(\alpha \,,\,\beta)$ plane allowed by triviality for (a) Model~I and (b)
Model~II. The remaining parameters take the following values: $M_H\!=\!400\;
{\rm GeV},\; M_h\!=\!100\;
{\rm GeV},\; \Mg\!=\!350\; {\rm GeV},\; M_t\!=\!150\; {\rm GeV},\;
\Ma\!=\!280\; {\rm GeV}$.
\end{enumerate}


\begin{thebibliography}{99}
\bibitem{lep} ALEPH Collaboration, \prp {\bf 216} (1992) 253;\\
DELPHI Collaboration, P.~Abreu et al., \np {\bf 373} (1992) 3;\\
L3 Collaboration, O.~Adriani et al., \pl {\bf 303} (1993) 391;\\
OPAL Collaboration, M.~Akrawy et al., \pl {\bf 253} (1991) 511.
\bibitem{trivt}M.~Aizenman, \prl {\bf 47} (1981) 1; \cmp {\bf 86} (1982)
1;\\
J.~Fr\"{o}hlich, \np{\bf 200} [FS4] (1982) 281;\\
A.~D.~Sokal, Ann. Inst. H. Poincar\'{e} {\bf A 37} (1982) 317.
\bibitem{dn} R.~Dashen and H.~Neuberger, \prl {\bf 50} (1983) 1897.
\bibitem{trivl} M.~L\"{u}scher and P.~Weisz, \np{\bf 318} (1989) 705;\\
J.~Kuti, L.~Lin and Y.~Shen, \prl {\bf 61} (1988) 678;\\
A.~Hasenfratz, K.~Jansen, C.~B.~Lang, T.~Neuhaus and H.~Yoneyama, \pl
{\bf 199} (1987) 531;\\
A.~Hasenfratz, K.~Jansen, J.~Jers\'{a}k, C.~B.~Lang, T.~Neuhaus and
H.~Yoneyama, \np{\bf 317} (1989) 81;\\
G.~Bhanot, K.~Bitar, U.~M.~Heller and H.~Neuberger, \np{\bf 353} (1991)
551.
\bibitem{smphen} For example, D.~Froidevaux, in {\it Proceedings of the
ECFA Large Hadron Collider Workshop}, Aachen 1990, (G.~Jarlskog and
D.~Rein, eds.), Vol.~II, p.~444, and references therein.
\bibitem{me} D.~Kominis and R.~S.~Chivukula, \pl {\bf 304} (1993) 152.
\bibitem{gem} GEM Technical Design Report; GEM TN-93-262, SSCL-SR-1219;
Submitted by the GEM Collaboration to the Superconducting Super Collider
Laboratory (April 30, 1993).
\bibitem{marciano} W.~J.~Marciano and F.~E.~Paige, \prl {\bf 66} (1991)
2433.
\bibitem{gunplb} J.~F.~Gunion, \pl {\bf 261} (1991) 510.
\bibitem{kz} Z.~Kunszt and F.~Zwirner, \np {\bf 385} (1992) 3.
\bibitem{mssm} V.~Barger, M.~S.~Berger, A.~L.~Stange and
R.~J.~N.~Phillips, \pr {\bf D 45} (1992) 4128;\\
J.~F.~Gunion, R.~Bork, H.~Haber and A.~Seiden, \pr {\bf D 46} (1992)
2040;\\
H.~Baer, M.~Bisset, D.~Dicus, C.~Kao and X.~Tata, \pr {\bf D 47} (1993)
1062.
\bibitem{gunorr} J.~F.~Gunion and L.~Orr, \pr {\bf D 46} (1992) 2052.
\bibitem{ghkao} J.~F.~Gunion, H.~Haber and C.~Kao, \pr {\bf D 46} (1992)
2907.
\bibitem{bcps} V.~Barger, K.~Cheung, R.~J.~N.~Phillips and A.~L.~Stange,
\pr {\bf D 46} (1992) 4914.
\bibitem{hawaii} H.~Baer, M.~Bisset, C.~Kao and X.~Tata, \pr {\bf D 46}
(1992) 1067.
\bibitem{gw} S.~L.~Glashow and S.~Weinberg, \pr {\bf D 15} (1977) 1958.
\bibitem{hhg}J.~F.~Gunion, H.~E.~Haber, G.~L.~Kane and S.~Dawson, The
Higgs hunter's guide, (Addison-Wesley, Reading, MA, 1990).
\bibitem{sher} M.~Sher, \prp {\bf 179} (1989) 273.
\bibitem{ineq} J.~F.~Gunion and H.~Haber, \np {\bf 272} (1986) 1;\\
S.~Bertolini, \np {\bf 272} (1986) 77.
\bibitem{top} C.~-P.~Yuan et al., preprint MSU 93/25 (hep-ph 9311226).
\bibitem{cleo} R.~Ammar et al., \prl {\bf 71} (1993) 674.
\bibitem{zbbbar} L.~Rolandi and R.~Tanaka, Proc. XXVI Int. Conf. on High
Energy Physics, Dallas, 1992.
\bibitem{buras} A.~J.~Buras, M.~Misiak, M.~M\"{u}nz and S.~Pokorski,
preprint MPI-Ph/93-77 (hep-ph 9311345).
\bibitem{park} G.~Park, preprint CTP-TAMU-69/93 (hep-ph 9311207).
\bibitem{tri} H.~M.~Georgi, S.~L.~Glashow, M.~E.~Machacek and
D.~V.~Nanopoulos, \prl {\bf 40} (1978) 692.
\bibitem{kaufman} R.~P.~Kauffman and W.~Schaffer, preprint BNL-49061
(hep-ph 9305279).
\bibitem{ehlq} E.~Eichten, I.~Hinchliffe, K.~Lane and C.~Quigg, \rmp
{\bf 56} (1984) 579; {\bf 58} (1986) 1065 (E).
\bibitem{dicusw} D.~Dicus and S.~Willenbrock, \pr {\bf D 39} (1989) 751.
\bibitem{kleisstir} R.~Kleiss and W.~J.~Stirling, \np {\bf 262} (1985)
235.
\bibitem{summpre} D.~J.~Summers, \pl {\bf 306} (1993) 129.
\bibitem{loopdecay} J.~F.~Gunion, G.~Gamberini and S.~F.~Novaes, \pr
{\bf D 38} (1988) 3481;\\
T.~J.~Weiler and T.-C.~Yuan, \np {\bf 318} (1989) 337.
\bibitem{kks} E.~Braaten and J.~Leveille, \pr {\bf D 22} (1980) 715;\\
M.~Drees and K.~Hikasa, \pl {\bf 240} (1990) 445; {\bf 262} (1991) 497
(E);\\
R.~Kleiss, Z.~Kunszt and W.~J.~Stirling, \pl {\bf 253} (1991) 269.
\bibitem{djouadiga} A.~Djouadi, M.~Spira and P.~M.~Zerwas, \pl {\bf 311}
(1993) 255.
\bibitem{djouadigl} A.~Djouadi, M.~Spira and P.~M.~Zerwas, \pl {\bf 264}
(1991) 440.
\bibitem{nlo} P.~Aurenche et al., \zp {\bf 29} (1985) 459;\\
H.~Baer and J.~Owens, \pl {\bf 205} (1988) 377;\\
B.~Bailey, J.~Owens and J.~Ohnemus, \pr {\bf D 46} (1992) 2018.
\bibitem{aurenche} P.~Aurenche et al., in {\it Proceedings of the ECFA
Large Hadron Collider Workshop}, Aachen 1990, (G.~Jarlskog and D.~Rein,
eds.), Vol.~II, p.~83.
\bibitem{seez} C.~Seez et al., in {\it Proceedings of the ECFA
Large Hadron Collider Workshop}, Aachen 1990, (G.~Jarlskog and D.~Rein,
eds.), Vol.~II, p.~474.
\bibitem{cms} CMS Collaboration, Expression of Interest, presented at
the General Meeting on LHC Physics and Detectors, Evian-les-Bains,
France, March 1992.
\bibitem{kun84} Z.~Kunszt, \np {\bf 247} (1984) 339.
\bibitem{kts} Z.~Kunszt, Z.~Tr\'{o}cs\'{a}nyi and W.~J.~Stirling, \pl
{\bf 271} (1991) 247.
\bibitem{bal} A.~Ballestrero and E.~Maina, \pl {\bf 268} (1991) 437.
\bibitem{sumplb} D.~J.~Summers, \pl {\bf 277} (1992) 366.
\bibitem{hmrsb} P.~N.~Harriman, A.~D.~Martin, W.~J.~Stirling and
R.~G.~Roberts, \pr {\bf D 42} (1990) 798.
\bibitem{AZh} H.~Baer, C.~Kao and X.~Tata, \pl {\bf 303} (1993) 284.
\end{thebibliography}
\end{document}